\documentclass[12pt,a4paper,notitlepage]{article}
\usepackage[russian, english]{babel}
\usepackage{indentfirst}
\usepackage{graphicx}
\usepackage{subfigure}
\usepackage{pstricks}
\usepackage[centerlast,small]{caption2}
\usepackage{hhline}
\usepackage{longtable}

\begin{document}

\title{SUNDMAN STABILITY OF NATURAL PLANET SATELLITES}

\author{Lukyanov L.G.,\thanks{luka@sai.msu.ru} \,
Uralskaya V.S.\thanks{ural@sai.msu.ru} \\
\bf Lomonosov Moscow State University \\\bf Sternberg Astronomical
Institute, Moscow, Russia}

\date{}

\maketitle

PACS №: 95.10.Ce

\begin{abstract}

The stability of the motion of the planet satellites is considered in
the model of the general three-body problem (Sun-planet-satellite).
"Sundman surfaces" are constructed, by means of which the concept "Sundman stability" is formulated. The comparison of the Sundman stability with the results of Golubev's $c^2h $ method and with the Hill's classical stability in the restricted three-body problem is performed. The constructed Sundman stability regions in the plane of the parameters "energy - moment of momentum" coincide with the analogous regions obtained by Golubev's method, with the value $(c^2h) _ {cr}$.

The construction of the Sundman surfaces in the three-dimensional space of the specially selected coordinates $xyR$ is carried out by means of the exact Sundman inequality in the general three-body problem. The determination of the singular points of surfaces, the regions of the possible motion and Sundman stability analysis are implemented. It is shown that the singular points of the Sundman surfaces in the coordinate space $xyR$ lie in different planes. Sundman stability of all known natural satellites of planets is investigated. It is shown that a number of the natural satellites, that are stable according to Hill and also some satellites that are stable according to Golubev's method are unstable in the sense of Sundman stability.

\end{abstract}


\section{INTRODUCTION}

The study of the motion stability of the planet satellites
has been usually performed by means of the Hill surfaces (Hill, 1878)
constructed either for the model of the restricted three-body
problem (Proskurin, 1950), or for the problem of Hill (Hagihara,
1952). Since Golubev (1967, 1968, 1985) who developed
the method it is sometimes referred to as $c^2h$ method
for the general three-body problem, the stability analysis of the motions is
usually carried out by Golubev's method. This method is based
on the famous Sundman inequality (Sundman, 1912).

The search for regions of stable motions in the general
three-body problem is divided into two tasks:

- determination of stability regions in the plane of two
free parameters --- the constant of the energy integral and constant
of the moment of momentum integral,

- determination of stable regions in the space of the coordinates used.

Actually Golubev (1967) carried out the complete solution of the
first task by the introduction of "the index of Hill stability"
$s=c^2h$, where $h$ is the energy constant, and $c$ is the moment
of momentum constant. Golubev showed that the curve $s=s_ {cr}$ is
the boundary of the stability region in the plane of $ch$, where the
value of $s_ {cr}$ is calculated at the Eulerian inner libration point
$L_2$ with the use of the known equation of fifth power as
some function of body masses.

The solution of the second problem is considerably more complex,
since it requires the construction of the Sundman surfaces
in the multidimensional space of the coordinates used.
For the solution of this problem Golubev (1968) applied
the simplified Sundman inequality, with the aid of which
the solution of problem leads to the construction of Hill curves
in the plane of two rectangular coordinates $x$ and $y$.

Some authors (Marchal, Saari, 1975), (Zare, 1976), {\bfseries
(Marchal, Bozis, 1982)}, (Marchal, 1990) have elaborated the
Golubev's results with the use of the simplified Sundman inequality.

The construction of the regions of possible motions
in the space of the selected coordinates with the use
of the exact Sundman inequality is hindered by the large number
of variables. Thus, in the relative or Jacobian coordinate
systems the number of the variables is six: three coordinates
of one body and three of the other. Therefore, the construction
of the Sundman surfaces can be carried out in the six-dimensional
space of these coordinates.

In the large series of works (Szebehely, Zare, 1977), (Walker,
Emslie, Roy, 1980), {\bfseries (Donnison, Williams, 1983)},
(Donnison, 2009), (Li, Fu, Sun, 2010), (Donnison, 2010) and other
authors the determination of the regions of the possible motions is
carried out in the six-dimensional space of the Keplerian elements
$a_1, a_2$, $e_1, e_2$, $i_1$  and  $i_2$ or in the four-dimensional
space $a_1, a_2$, $e_1$ and $e_2$ for the planar problem.

At the same time, the construction of the Sundman surfaces and, thus,
the determination of the regions of possible motions and stability
regions can be conducted in the space of three variables.
This substantially facilitates the readability of results
and eases their application. This possibility is explained by the fact
that the exact Sundman inequality depends on the coordinates only
by means of three quantities --- three mutual distances between the bodies.

The article (Lukyanov and Shirmin, 2007) is likely to be the first
work that confirmed this possibility. In this work the mutual
distances between the bodies are the rectangular coordinates. The
existence of the Hill surfaces's analogue for the general three-body
problem in the space of the mutual distances is shown here using the
exact Sundman inequality. The stability regions are determined in
the space of the mutual distances and have the form of an infinite
"tripod".

Lukyanov (2011) used another choice of three coordinates. The coordinate system
is determined by the accompanying triangle of mutual positions of bodies. Namely,
the origin of the coordinate system coincides with one of the bodies, the axis $x$
is directed towards the second body, the axis $y$ is perpendicular to the axis $x$
and lies in the plane of the triangle. In this coordinate system the position of all bodies
is determined by three coordinates $x, y, R$, where $x$ and $y$ are the coordinates
of the third body, and $R$ is the distance between the first and second bodies.
The system of coordinates $xyR$ is to a certain degree similar to the rotating coordinate
system in the restricted three-body problem following the motion of the basic bodies.

Then, in the space of the coordinates $xyR$ the Sundman surfaces are
constructed, the singular points of surfaces (coinciding with the
Euler and Lagrange libration points) are located, the regions of
the possible motions and Sundman stability regions are determined.
The stability region of any body relative to the other body in the space
$xyR$ has the form, similar to an infinite "spindle". No restriction on
the masses of bodies or their mutual positions is assumed in this
case.

In the present work the method of constructing the regions of the
possible motions (Sundman lobes) in the general three-body problem
is presented. The Sundman stability analysis of the natural
satellites of the planets is carried out, using the high-precision
ephemeris of the natural satellites of planets calculated on the
web-site "Natural Satellite Data Center" (NSDC) (Sternberg
Astronomical Institute, Moscow)

\noindent $(http://www.sai.msu.ru/neb/nss/index.htm)$.

\section{SUNDMAN SURFACES}

The regions of possible motions of bodies in the
general three-body problem are determined by Sundman inequality
\begin{equation}\label{1}
 (U-C)J\geq B,
\end{equation}
where the force function $U$ and the barycentric moment of
inertia $J$ are determined by the expressions:
\begin{equation}\label{2}
U=\frac{Gm_1m_2}{R_{12}}+\frac{Gm_2m_3}{R_{23}}
+\frac{Gm_3m_1}{R_{31}},
\end{equation}
\begin{equation}
\label{3}
J=\frac{m_1m_2R_{12}^2 + m_2m_3R_{23}^2 + m_3m_1R_{31}^2}{m},
\end{equation}
$C=-h$ is the analogue of the Jacobi constant, $h$ is the
energy constant, $B=c^2/2$ is the Sundman constant, $c$ is the
constant of the integral of area.

Here: $G$ is the universal gravitational constant, $m_1, m_2, m_3$ are
the masses of bodies, $m=m_1+m_2+m_3$ is the total mass of the
system, $R_{12}$, $R_{23}$, $R_{31}$ are the mutual distances
between the bodies.

Constants $C$ and $B$ are determined by the initial conditions in
the barycentric coordinate system from the relationships:
\begin{equation}\label{2a}
\begin{array}{c}
\displaystyle C=-h=U-m_1\frac{V_1^2}{2} -
m_2\frac{V_2^2}{2} -m_3\frac{V_3^2}{2}, \\
\displaystyle B=\frac{c^2}{2} = \frac{1}{2}(m_1\mathbf{r}_1 \times
\mathbf{V}_1 + m_2\mathbf{r}_2 \times \mathbf{V}_2 +
m_3\mathbf{r}_3 \times \mathbf{V}_3)^2, \\
\end{array}
\end{equation}
$\mathbf{r}_i,\, \mathbf{V}_i, \, (i=1, 2, 3)$ are the barycentric
state and speed vectors of the bodies.

The boundary of the region of possible motions can be
established if in (\ref{1}) inequality is replaced with
equality:
\begin{equation}
\label{4}
(U- C) J = B.
\end{equation}

This equality determines the equation of the Sundman surface. In the general
case the mutual distances in (\ref{4}) depend
on nine coordinates of three moving bodies, which substantially
hampers the construction of the Sundman surfaces. The transformation to the
relative coordinate system makes it possible to reduce the number of
coordinates to six. However, in this case the construction of the
Sundman surfaces should be conducted in the
six-dimensional space. Furthermore, the number of coordinates can be
reduced to three if the positions of bodies are determined by the
following special coordinates.

The position of the body $M_2$ relative to the body $M_1$ will be
characterized by the abscissa $R$ on the axis $M_1X$. we will define
the position of the body $M_3$ relative to $M_1$ by the
rectangular coordinates $X$ and $Y$ in the system of $M_1XY$, which
always lies in the plane that passes through all three bodies. The
positions of the bodies in the coordinate system $M_1XYR$ are defined by
three quantities - coordinates $X$, $Y$ and $R$, which allows us to
construct the Sundman surfaces in the three-dimensional space.

We will use a dimensionless
system of coordinates $M_1xy$ making the substitution
\begin{equation}
\label{6} X=Rx, \quad Y=Ry.
\end{equation}

Then the mutual distances between the bodies can be expressed
in terms of three quantities $x, y, R$. The Sundman surface equation transforms
to the form of the functions of three variables
\begin{equation}\label{7}
\begin{array}{c}
\displaystyle S(x, y, R) = (U-C)J = \left[\frac{G}{R}\left(m_1m_2 +
\frac{m_2m_3}{\sqrt{(x-1)^2+y^2}} +
\frac{m_3m_1}{\sqrt{x^2+y^2}}\right)-C\right]\times\\[16pt]
\displaystyle \times\frac{R^2}{m}\left\{m_1m_2 + m_2m_3
\left[(x-1)^2+y^2\right] + m_3m_1\left(x^2+y^2\right)\right\} =
\frac{c^2}{2} = B.
\end{array}
\end{equation}

Equation (\ref{7}) allows us to conduct the
construction of the Sundman surface in the three-dimensional
cartesian space of variables $xyR$.

The singular points of the Sundman surfaces are
determined from the system of three algebraic equations
\begin{eqnarray}\label{9}
\nonumber \frac{\partial S}{\partial x} &=&
\frac{2R^2(U-C)}{m}[m_2m_3(x-1)+m_3m_1x]-\frac{G J}{R}
\left[\frac{m_2m_3
(x-1)}{r_{23}^3}+\frac{m_3m_1x}{r_{31}^3}\right]=0, \\
\frac{\partial S}{\partial y} &=& y\left[\frac{2R^2(U-C)}{m}(m_2
m_3+m_3m_1)-\frac{G J}{R}\left(\frac{m_2m_3}{r_{23}^3}+
\frac{m_3m_1}{r_{31}^3}\right)\right] =0, \\
\nonumber \frac{\partial S}{\partial R}&=& \frac{J(U-2C)}{R}=0.
\end{eqnarray}

From the third equation of this system it is possible to determine
the mutual distance $R$ between the bodies $M_1$ and $M_2$
in the form of the function of unknowns $x$, $y$ and constant $C$:
\begin{equation}\label{10}
R=\frac{G}{2C}\left(m_1m_2+\frac{m_2m_3}{\sqrt{(x-1)^2 + y^2}} +
\frac{m_3m_1}{\sqrt{x^2+y^2}}\right).
\end{equation}

Substituting this expression for $R$ to the first two equations of
set (\ref{9}), we will obtain the system of two equations with two
unknowns $x$ and $y$:
\begin{equation}\label{11}
\begin{array}{rc}
\displaystyle\left(m_1m_2+\frac{m_2m_3}{r_{23}} +
\frac{m_3m_1}{r_{31}}\right)[m_2m_3(x-1)+m_3m_1x]- &\\[12pt]
\displaystyle - \left[\frac{m_2m_3
(x-1)}{r_{23}^3}+\frac{m_3m_1x}{r_{31}^3}\right](m_1m_2 + m_2m_3
r_{23}^2 + m_3m_1r_{31}^2)&=0, \\[12pt]
\displaystyle y\left[\left(m_1m_2+\frac{m_2m_3}{r_{23}} +
\frac{m_3m_1}{r_{31}}\right)(m_2 m_3+m_3m_1)\right. - &\\[12pt]
\displaystyle - \left.\left(\frac{m_2m_3}{r_{23}^3}+
\frac{m_3m_1}{r_{31}^3}\right) (m_1m_2 + m_2m_3
r_{23}^2 + m_3m_1r_{31}^2)\right]&=0. \\
\end{array}
\end{equation}

The second equation in set (\ref{11}) can be satisfied in two
ways: to set $y=0$ or to consider as zero the entire coefficient
in the brackets with $y$. The first possibility ($y=0$) leads to the
collinear singular points, the second --- to the triangular.

For $y=0$ we obtain from set (\ref{11}) one equation for
the determination of the coordinate $x$ of the collinear singular
points
\begin{equation}\label{12}
\begin{array}{c}
\varphi (x) \displaystyle
=\left(m_1m_2+\frac{m_2m_3}{\sqrt{(x-1)^2 }} +
\frac{m_3m_1}{\sqrt{x^2}}\right)\left[m_2m_3(x-1)+m_3m_1x\right] - \\
- \displaystyle \left[\frac{m_2m_3}
{(x-1)\sqrt{(x-1)^2}}+\frac{m_3m_1}{x\sqrt{x^2}}\right] [m_1m_2 +
m_2m_3(x-1)^2 + m_3m_1x^2] = 0.\\
\end{array}
\end{equation}

The derivative $\varphi'(x)$ is always positive, and the following
limits take place:
\begin{equation}\label{13}
\lim\limits_{x \to
\mp\infty} \varphi (x)=\mp\infty, \quad \lim\limits_{x \to 0\mp0}
\varphi (x)=\pm\infty, \quad \lim\limits_{x \to 1\mp0} \varphi
(x)=\pm\infty.
\end{equation}

This proves the existence of three real solutions of the equation
$\varphi(x)=0$, which, in their turn, determine three collinear singular
points of the family of the Sundman surfaces in space $xyR$.

\begin{equation}\label{14}
L_i= \left(x_i,\, 0,\,
\frac{Gm_1m_2}{2C}+\frac{Gm_2m_3}{2C\sqrt{(x_i-1)^2}} +
\frac{Gm_3m_1}{2C\sqrt{x_i^2}}\right), \;(i=1, 2, 3),
\end{equation}
where the coordinates $x_i$ are determined by the numerical
solution of equation (\ref{12}).

But if $y\neq 0$, then after simple conversions we obtain two
triangular solutions of set (\ref{11})
\begin{equation}\label{l5}
 R_{23}=R_{31}=R,
 \end{equation}
which determine two triangular singular points in the space $xyR$:
\begin{equation}\label{16}
L_{4,5}= \left(\frac{1}{2},\, \pm\frac{\sqrt{3}}{2},\,
G\,\frac{m_1m_2+m_2m_3+m_3m_1}{2C}\right).
\end{equation}

The obtained collinear and triangular singular points correspond to
the collinear Euler and triangular Lagrange solutions known in the
general three-body problem.

Collinear singular points in the space
$xyR$ lie in different planes $R=R_i$, i.e.,  $R_i\neq
R_j$, and for triangular singular points the following equality is
fulfilled: $R_4=R_5$.

\begin{figure}[t]
\centering
\includegraphics[totalheight=12cm]{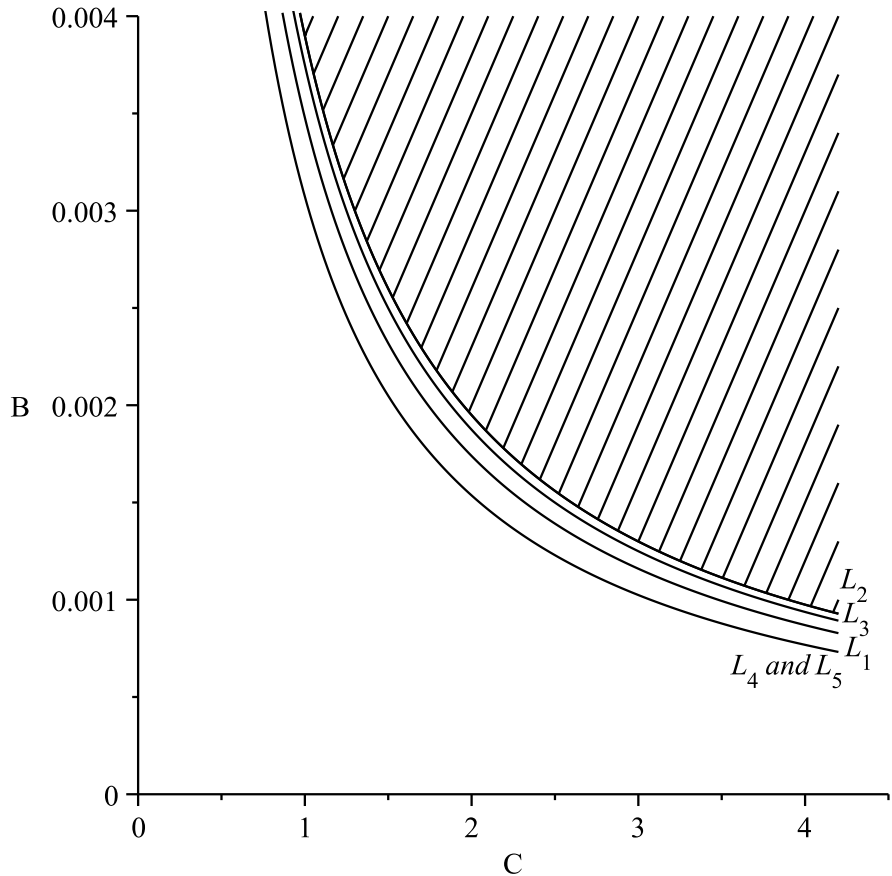}
\caption{The stable region (hatched) and the boundaries $(L_1, L_2,
L_3, L_4\, and \,L_5)$ of different topological types of the regions
of the possible motion for $\displaystyle
\frac{m_1}{9}=\frac{m_2}{3}=\frac{m_3}{1}$. } \label{Fig0}
\end{figure}

Knowing the coordinates of singular points and constant $C$, from
formula (\ref{7}) the values of Sundman constant $B_1, B_2, B_3$ and
$B_ {4,5}$ in all singular points $L_i, (i=1,2,\dots,5)$ are
calculated. Constants $C$ and $B_i$ are connected by reciprocal
proportion
\begin{equation} \label{17}
B_i=\frac{G^2}{4mC}\Bigl[m_1m_2+m_2m_3(r_{23}^2)_i +
m_3m_1(r_{31}^2)_i\Bigl]
 \left[m_1m_2+\frac{m_2m_3}{(r_{23})_i}+\frac{m_3m_1}{(r_{31})_i}
  \right]^2,
\end{equation}
where $(r_{23})_i$ and $(r_{31})_i$ are calculated at the
singular point $L_i$.

The relations (\ref{17}) have been established by Golubev (1967) in the $c^2h $
method.

\begin{figure}[t]
\centering
\includegraphics[totalheight=14cm]{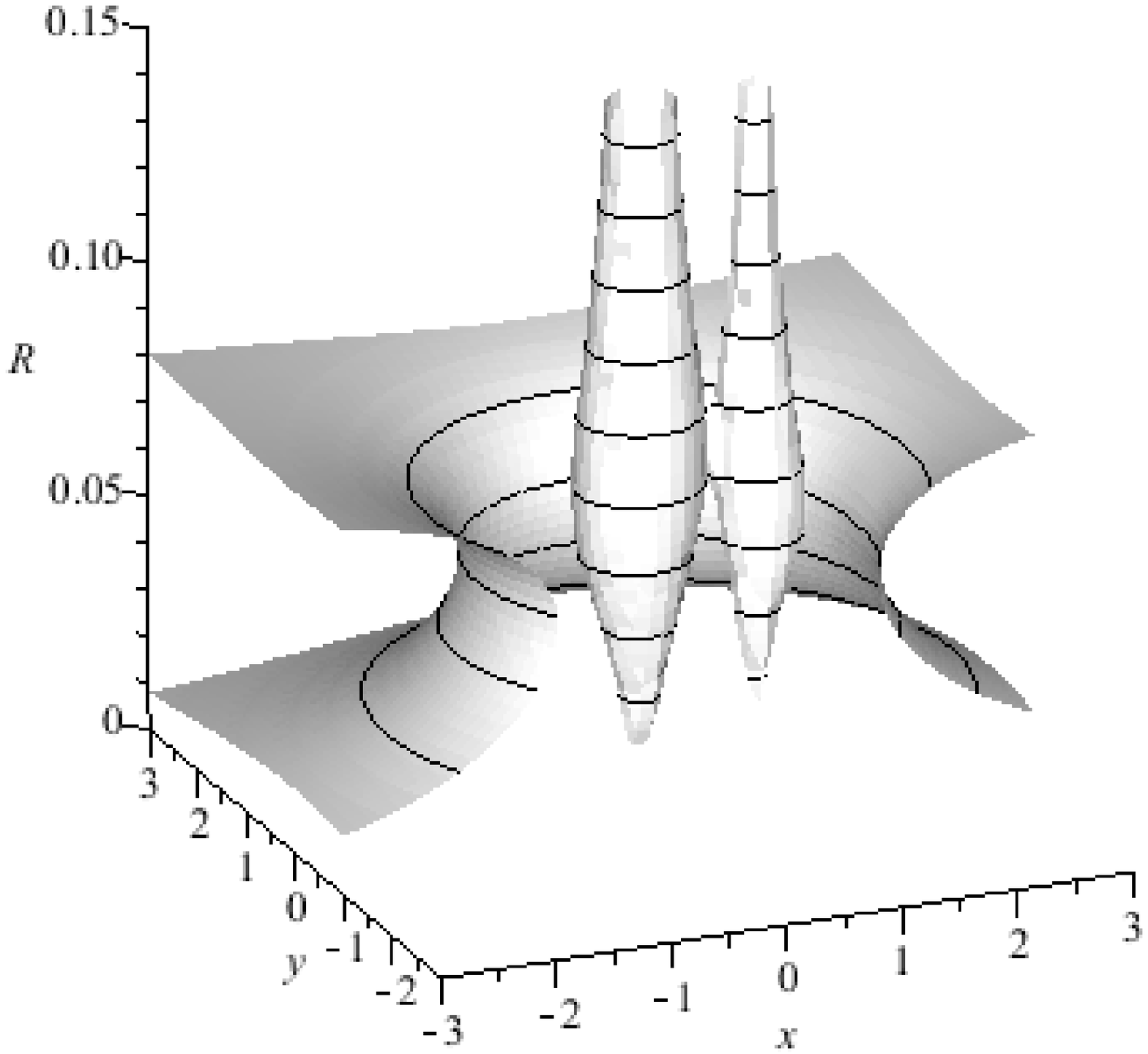}
\caption{General view of Sundman surfaces for $\displaystyle
\frac{m_1}{9}=\frac{m_2}{3}=\frac{m_3}{1}$, $C$=2 and
$B=B_2$. Surfaces are represented in the field of the space limited
to planes: $x =-3, \, x=3, \, y =-1, \, y=3, \, R=0, \, R=0.7$.}
\label{Fig1}
\end{figure}

\begin{figure}[t]
\centering
\includegraphics[totalheight=7.5cm]{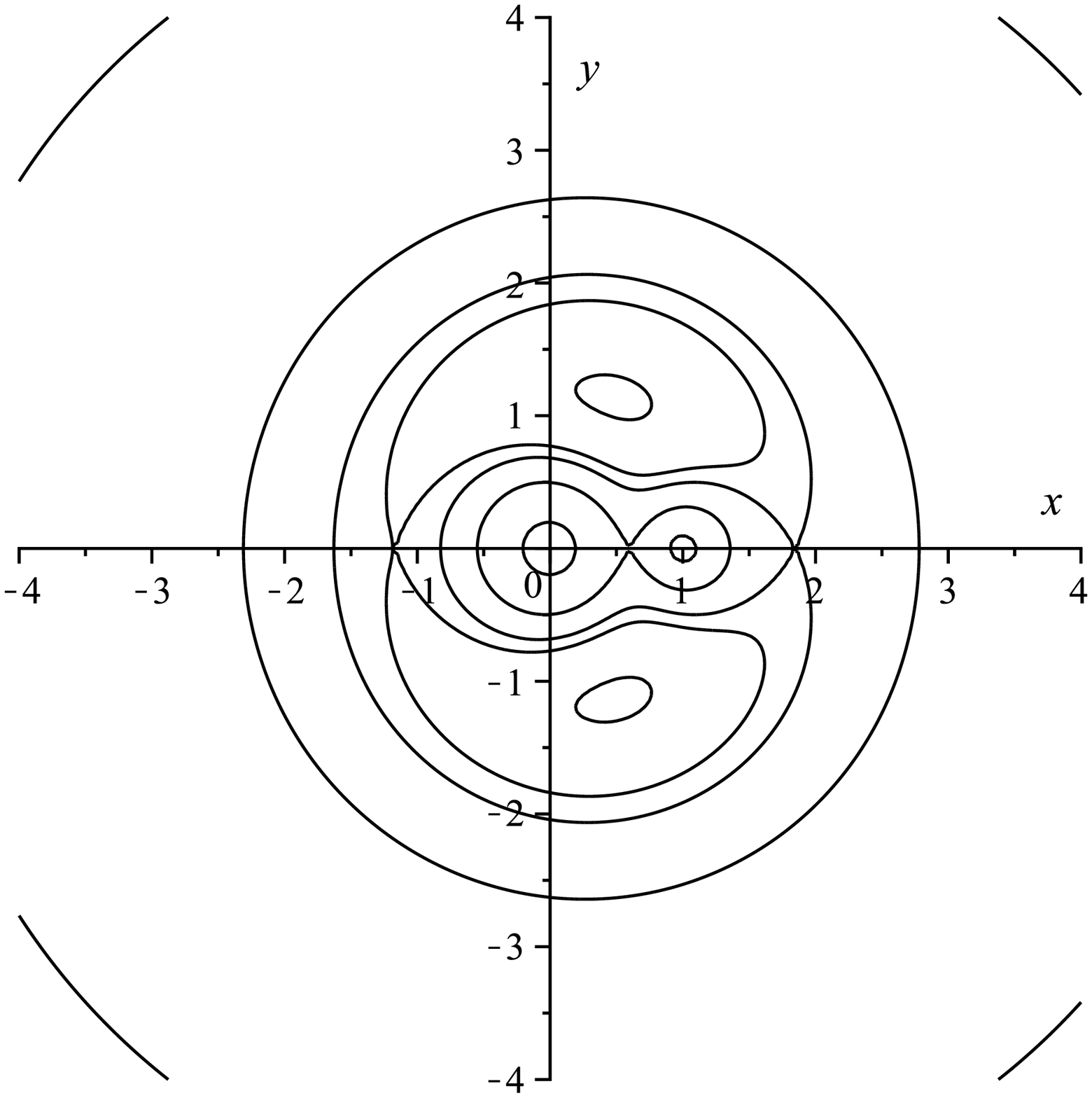}
\includegraphics[totalheight=7.5cm]{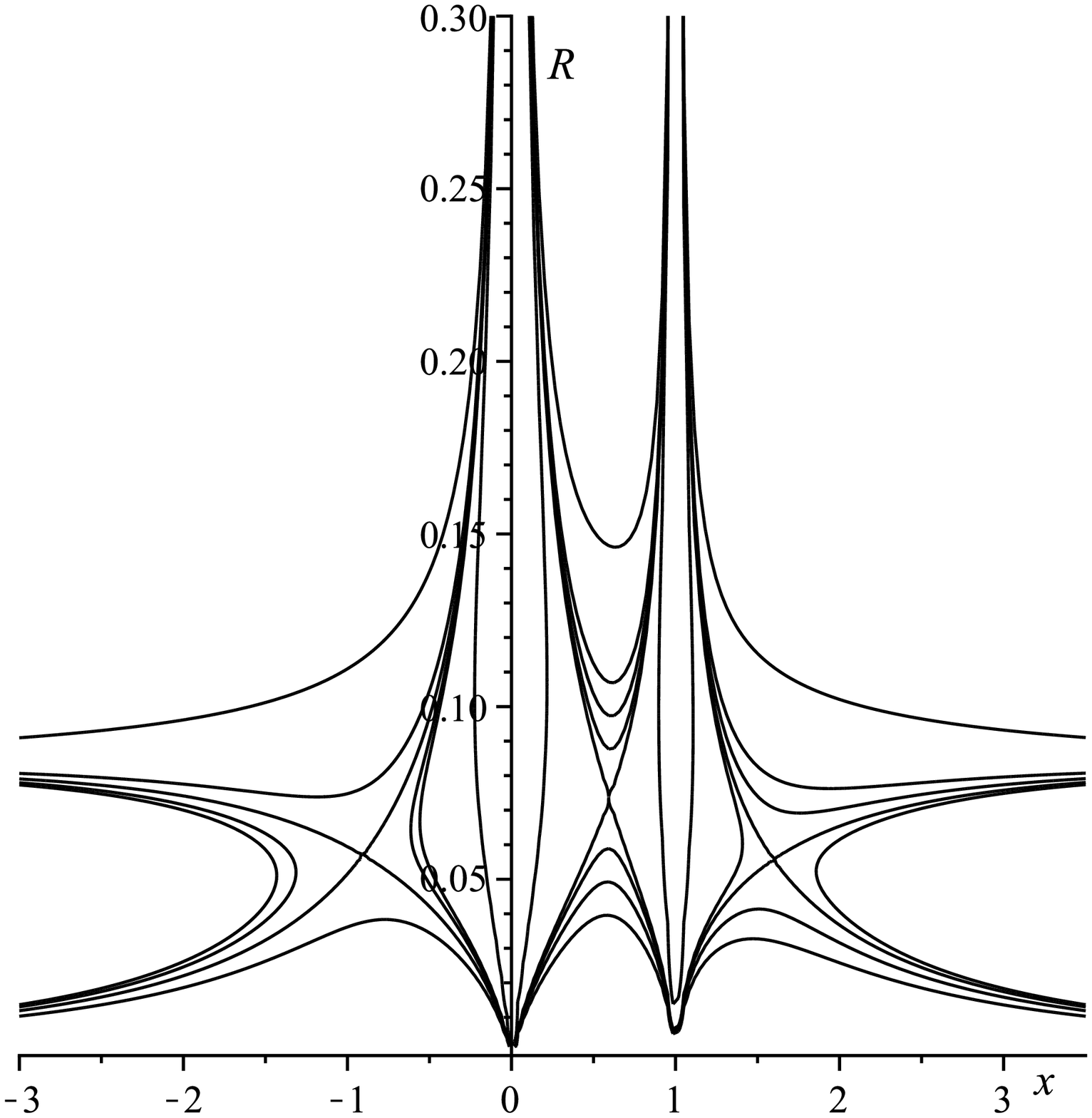}\vskip-0.25cm
\caption{The sections of the Sundman surfaces for $\displaystyle
\frac{m_1}{9}=\frac{m_2}{3}=\frac{m_3}{1}$ by the plane $R=R_2$
(left), by the plane $y=0$ (right).} \label{Fig2}
\end{figure}

Singular points are the points of bifurcation, in which a
qualitative change in the shape of Sundman surface occurs. The
curves (\ref{17}) on plane $CB $ are the boundaries of the
topologically different regions of the possible motion. The curve
$L_2$ limits the Sundman stable region of the body $M_3$, and this
stable region is shaded (Fig.\ref{Fig0}).

The general form of the Sundman surfaces for three bodies with the
mass ratio proportional to 9:3:1 is shown in Fig.\ref{Fig1}; the
section of the Sundman surfaces by the planes $R= R_2$ and $y=0$ is
presented in Fig.\ref{Fig2}.

In the general three-body problem, as in the restricted problem, the
concept of Hill stability is conserved. But, to distinguish it from the
restricted problem in the general three-body problem, we will call
this stability \emph {Sundman stability}.

We will call the motion of the body $M_3$ in the general three-body problem
stable on Sundman if there are such regions of the
possible motions, limited by the appropriate Sundman surfaces,
inside which the body $M_3$ will be always (at any instant of time)
located at a finite distance from one of the bodies $M_1$ or $M_2$.
In other words, the body $M_3$ will be an eternal satellite of one
of $M_1$ or $M_2$ bodies, while bodies $M_1$ or $M_2$ can be at any
distance one from the other, including infinite.

Criterion of Sundman stability is the inequality
\begin {equation} \label{18}
B\geq B_2,
\end {equation}
where $B_2$ is the value of Sundman constant in the inner Euler
libration point $L_2$. The fulfillment of this condition guarantees
that the body $M_3$ can be: in some "spindly" surfaces (see Fig.2,
3) remaining the eternal satellite of a body $M_1$; or in other
"spindly" surfaces, remaining the satellite of body $M_2$, or in a
remote open oval area, when the distance between bodies $M_1$ and
$M_2$ remains finite, not exceeding $Gm_1m_2/C $. This last case can
be treated as Sundman stability of the relative motion of bodies
$M_1$ and $M_2$.

Thus, for (\ref{18}) any pair of bodies
will have Sundman stability if at the initial instant the bodies forming this
pair  are in one of these regions of
stability. The loss of
stability (body $M_3$ leaving the
"spindly" area) occurs if the value $R $ is
close enough to its value $R_2$ at the libration point $L_2$.

\section{SUNDMAN STABILITY \\ OF THE PLANET SATELLITES' MOTION}

The analysis of Sundman stability of the motion of all known natural
planet satellites of the Solar System is investigated with the
presented theory. The ephemeris of all planet satellites are
calculated with the most uptodate theories implemented on the NSDC
web-site (Natural Satellites Data Center) (Sternberg Astronomical
Institute, Moscow, Russia), constructed by Emelyanov, Arlo (2008)
$(http://www.sai.msu.ru/neb/nss/index.htm)$. From these ephemeris
constants $C$, $B$ and $B_2$ were calculated in the barycentric
coordinate system. Sundman stability was determined from formula
(\ref{18}).

\begin{figure}[t]
\centering
\includegraphics [totalheight=7.cm] {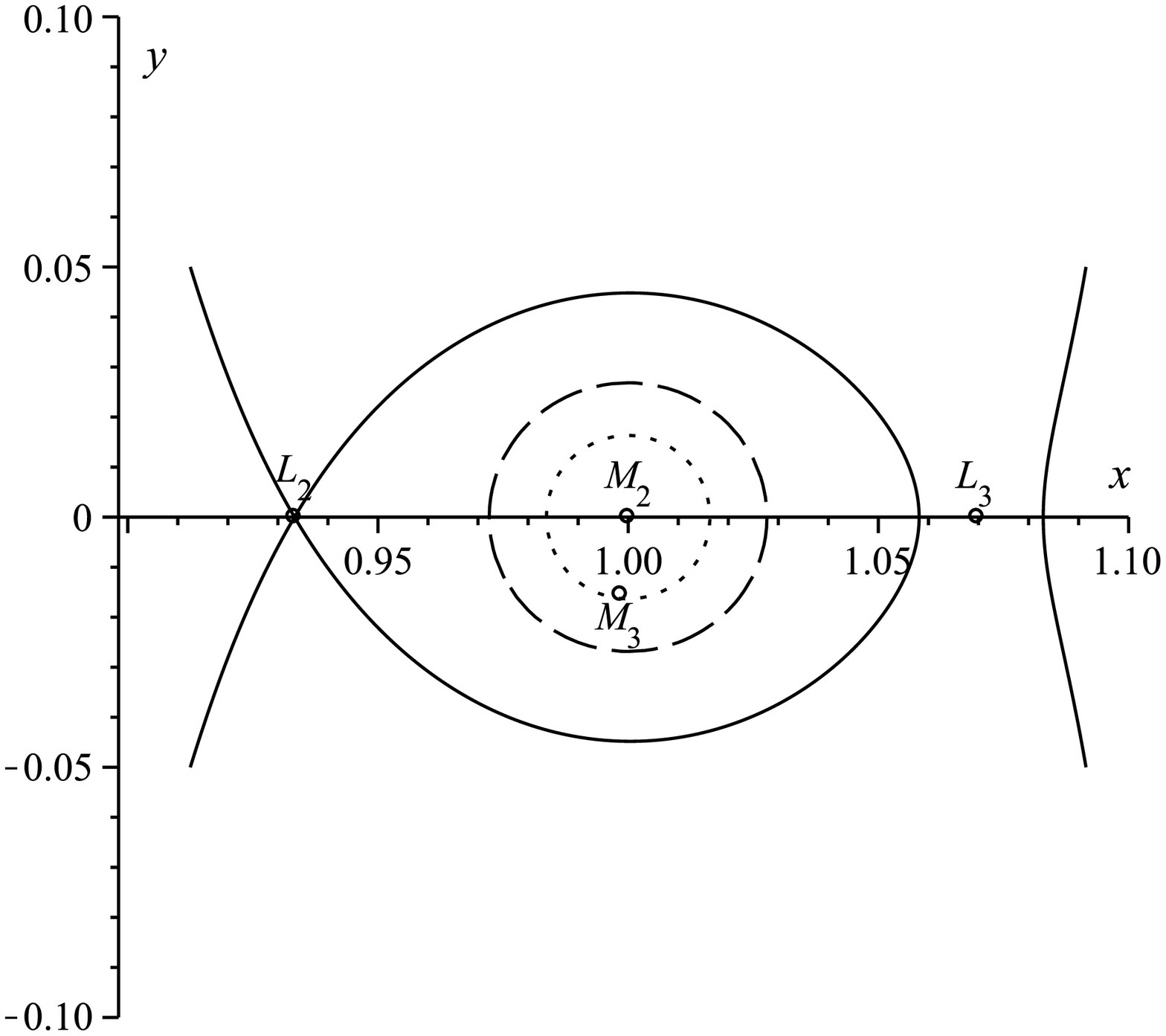}
\includegraphics [totalheight=7.cm] {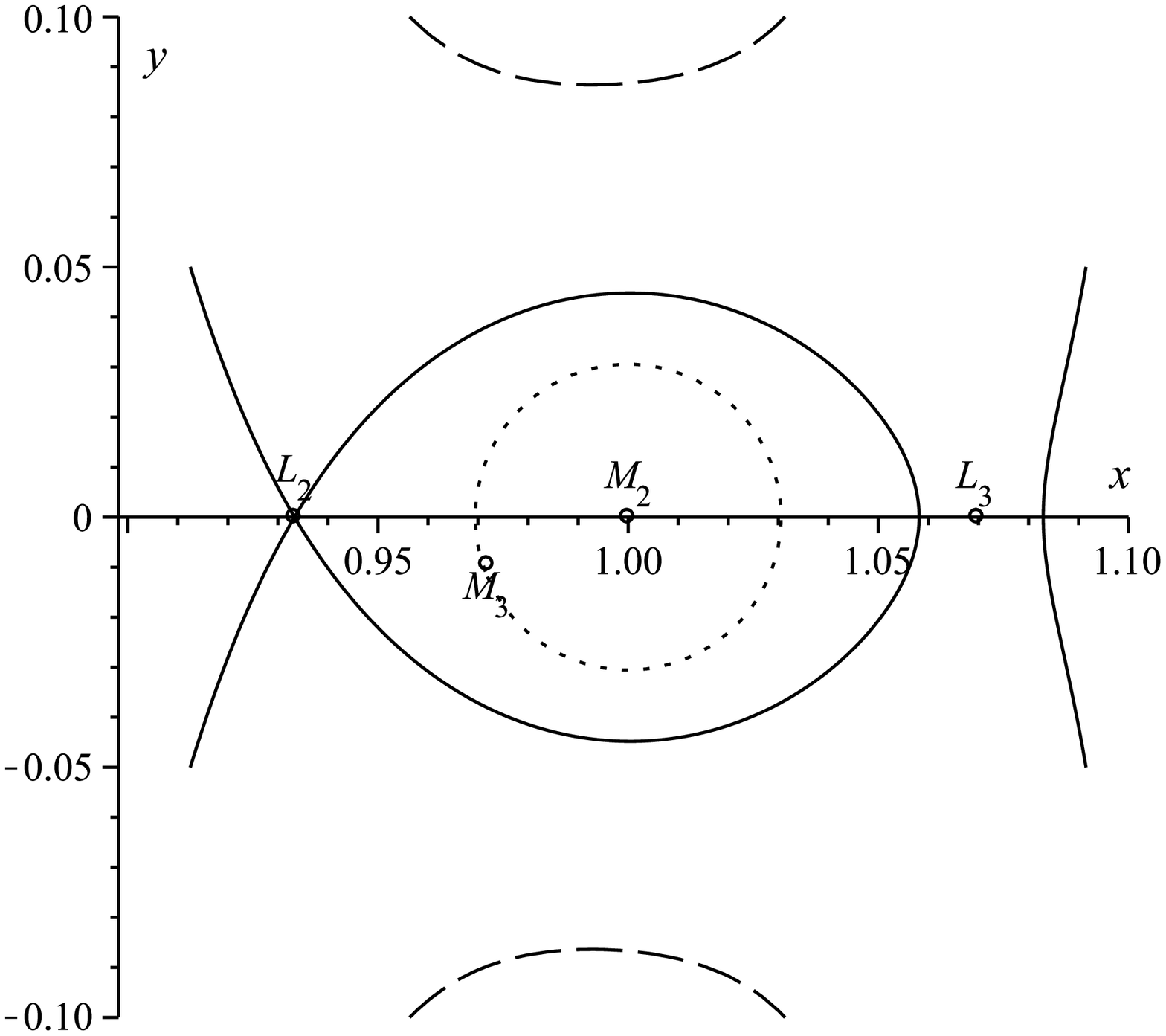}
\centering \caption {The sections of the Sundman surfaces by the
plane of $R=R_2$ for Jovian satellites J6 Himalia (left) and J9
Sinope (right). Here: the solid line $B_2$ is the boundary of
Sundman stability region, the dash line is satellite's Sundman
curve, the dotted line is approximate region of satellite motions.}
\label{fig3}
\end{figure}

For each satellite the construction of the Sundman
surface sections by the coordinate plane $xy$ was also conducted. The
sections are given for the Jovian satellite J6 Himalia and J9 Sinope
(Fig.\ref{fig3}). Himalia's Sundman curve is within the Sundman
stability region, Sinope's curve is outside. In spite of the location
of the Sinope orbit inside the Sundman lobe, which corresponds to
Sundman constant value $B=B_2$, its energy is sufficiently high, and
it has a potential capability to leave this lobe (see the dash
curve). But this does not mean that the satellite will leave the
vicinity of the planet without fail. Sundman instability means that
the Sundman surfaces are open and allow it to leave the vicinity of
planet. But the Sundman surfaces do not tell if this will occur or
not. The same case is for the Hill stability.

Lukyanov (2011) showed the Sundman stability of the Moon motion. The
Sundman stability results for the rest satellites are given in Tables
(\ref{Tab0}-\ref{Tab4}). The tables also list the results of the
classical Hill stability. All satellites are located in the
order of increasing semimajor axes of their orbits around the planet.
The relative masses of distant planet satellites obtained from
satellite photometric observations (Emelyanov, Uralskaya, 2011), are
taken from the web-site NSDC
(http://www.sai.msu.ru/neb/nss/index.htm).

The Martian satellites (Phobos and Deimos) have Hill and Sundman
stability (Tab.\ref{Tab0}).

The main and the inner satellites of Jupiter have Hill stability and
Sundman stability and are not included in the tables. The distant
satellites, which have prograde and retrograde orbits, are of
special interest. All prograde satellites of Jupiter have Hill and
Sundman stability (Tab.\ref{Tab1}). All retrograde satellites with
$a > 18.34\cdot10^{6}\, km$ are unstable according to Hill and
Sundman, independently from their masses. An exception is the
satellite S/2003 J12 ($a = 19\cdot10^{6}\, km$, $i = 145.8^{o}$, $e
= 0.376$) with a relatively small mass, which has Hill stability and
Sundman stability.

The situation is different for the satellites of Saturn. The main,
inner and distant prograde satellites of Saturn, which belong to the
Gallic group ($i = 34^{o}$) and Inuit group ($i = 45^{o}$), have
Hill stability and Sundman stability (Tab. \ref{Tab2}). The
retrograde satellites with $a < 18.6\cdot10^{6}\, km$ have Hill and
Sundman stability, with $a > 18.6\cdot10^{6}\, km$ have Hill
stability, but Sundman instability. Furthermore, Sundman unstable is
also the satellite S LI Greip with the semimajor axis $a =
18.1\cdot10^{6}\, km$.

The main and inner satellites of Uranus have the Hill stability and
Sundman stability. The stability results coincide for all distant
satellites, except for the most distant satellite U XXIV Ferdinand
($a = 20.9\cdot10^{6}\, km$), which has Sundman instability
(Tab.\ref{Tab3}).

Triton and the Neptune inner satellites have the Hill and Sundman
stability. Two distant Neptune satellites have Sundman instability:
N X Psamathe ($a = 46\cdot10^{6}\, km$)  and N XIII Neso ($a =
48\cdot10^{6}\, km$). The rest of Neptune satellites have Hill
stability and Sundman stability (Tab.\ref{Tab4}).

The comparison of the results of the Sundman stability and Hill
stability shows that Hill stability always follows from the Sundman
stability, but the reverse assertion is not correct. It is caused by
the fact that, in contrast to Hill's model, in the Sundman model the
satellite masses are not zero, but are finite. Therefore,
each satellite of any planet has an individual value of the Sundman
constant $B_2$, while in Hill's model all satellites of any planet
have the same value of the Hill constant $C_2$.

The comparison of the obtained results with Golubev $c^2h$ method
is carried out in two directions:

--- comparison of the stability criteria used,

--- comparison of the obtained regions of the possible motions.

Analytical forms of stability criterion in our work $B\geq B_2$ and
in Golubev's method $c^2h\leq(c^2h)_{cr}$, are the same. But the
calculation of the constants in the left and right sides of the
inequalities is carried out using different formulas. This leads to
some differences in numerical results. The comparison with the
results of the work (Walker et al, 1980) for the satellites J1-J13
shows that the Sundman stability or instability of the these
satellites, obtained in our work, agrees with the results of Walker et al.
(1980) for all satellites, except for four satellites of Jupiter with
retrograde motion, J VIII, J IX, J XI and J XII. For these satellites
we obtained instability, while in the work cited these satellites
were indicated as being stable. This is likely to be due to the
approximation of the three-body problem by two problem of two bodies
and also by the neglect of orbit inclinations.

We conducted the construction of regions of possible motions
in the three-dimensional space of $xyR$, while in all works of other
authors the value of $R$ is excluded from the examination, and the
constructions of regions of possible motions are conducted in
the $xy$-plane. For this reason in the $c^2h$ method it is not
possible to get a number of important results. For
example, it cannot be obtained that the loss of stability
(withdrawal of the body $M_3$ from the stability region) can
occur only for a certain distance between the bodies $M_1$ and
$M_2$. Generally, Sundman curves in the plane $R= \mathrm {const}$
with a change in $R$ can sharply and qualitatively differ from
Hill's curves, as shown by Lukyanov (2011).

\section{DISCUSSION}
The famous Sundman inequality in the general three-body problem
takes the form
\begin{equation}\label{D1}
(U-C)J- B \geq \frac{\dot{J}^2}{8}.
\end{equation}
For the material motions of bodies, i.e., with the fulfillment
of conditions $\dot{J}^2\geq 0$, it determines the regions
of possible motions satisfying the inequality
\begin{equation}\label{D2}
(U-C)J\geq B.
\end{equation}

The boundaries of the region of possible motions
are determined by the equation
\begin{equation}\label{D3}
(U-C)J= B,
\end{equation}
which we call the equation of the \emph {Sundman surface}, while
the stability in Hill's sense for the three-body problem
--- \emph {Sundman stability}. By analogy with the surfaces of
the zero speed in the restricted three-body problem, we may
call the Sundman
surfaces in the general three-body problem
\emph {the surfaces of zero rate of change of
the barycentric moment of inertia of bodies} $ (\dot {J} =0) $.

The determination of the Sundman stability and the construction of
the Sundman curves in the plane of parameters $C$ and $B$ (see Fig.
1) is completely solved by Golubev\footnote {in the English-language
literature the surname Golubev is frequently written incorrectly.}
(1967) in his $c^2h$ method (in our designations $c^2=2B, h=-C$).
Now this method is called Golubev's method.

Golubev's method determines not the surfaces, but the Sundman curves
located in the plane of the triangle formed by the mutual
distances between the bodies. The mutual distances
between the bodies $R_ {13}$ and $R_ {23}$ are substituted
by the relative values $R_ {13} /R$ and $R_ {23} /R$,
and the value of $R=R_ {12}$ is generally excluded from examination.

Equation of "current" Sundman curve in Golubev's method has the form of
the hyperbola $CB= \mathrm {const}$. If in this case the constants
$C$ and $B$ are expressed in term of any other variables, then, in its
turn, the task of construction of the Sundman curves in the space of
these variables arises. Thus, in the large series of works of
(Szebehely and Zare, 1976), (Walker, 1983), (Donnison, 2010) and
many other authors the task of constructing Hill-Sundman curves
and determination of stability regions in the general three-body
problem is solved by Golubev's method in the space of six quantities:
semimajor axes $a_1, a_2$, eccentricities $e_1, e_2$ and
inclinations $i_1, i_2$, for calculation of the constants $C$ and
$B$ the approximation of three-body problem by two problems of two
bodies is used. This introduces a certain error to the solution of
problem. Besides, the value of $R$ remains unknown.

For the representation of the Sundman curves on the plane $xy$,
Golubev (1968) considered another method. He used the simplified
Sundman inequality instead of the exact inequality (\ref{D2})

\begin{equation}\label{D4}
U^2J\geq BC,
\end{equation}
which is the consequence of inequality (\ref{D2}) and is
obtained after the multiplication of inequality (\ref{D2}) by $U$,
taking into account inequalities $C> 0$ and $U> C$. Inequality
(\ref{D4}) does not reflect the entire diversity of the Sundman
surfaces.

Like the $c^2h$ criterion (obtained from the condition of the
positivity of the discriminant of the quadratic trinomial for $R$
from the left side of the Sundman inequality), simplified inequality
(\ref{D4}) does not contain the mutual distance $R_ {12} =R$.
Therefore, by means of inequality (\ref{D4}), it is possible to
construct not the surfaces, but the Sundman curves in the plane of
relative coordinates $xy$. The construction of these curves was
subsequently conducted in the works of (Marchal, Saari, 1975),
{\bfseries (Marchal, Bozis, 1982)} and other authors.

Thus, the task of constructing the Sundman surfaces
in the space of the coordinates used remained
incomplete before the publication
(Lukyanov, Shirmin, 2007) and (Lukyanov, 2011) appeared.
Lukyanov, Shirmin (2007) used the mutual distances
between the bodies as the coordinates. This made possible to construct
exact Sundman surfaces in the three-dimensional space
of mutual distances. Lukyanov (2011) used the more convenient rectangular coordinate
system $xyR$, determined by the accompanying triangle
of mutual positions of three bodies.

In these works the exact Sundman inequality (\ref{D2}) is used and,
therefore, the value of $R$ is not excluded from the examination.
In this case no simplifications or assumptions are applied. The
construction of the Sundman surfaces is implemented in the
three-dimensional space of the coordinates used with the
determination of the singular points of surfaces, regions of the
possible motion and Sundman stability regions.

Regions of the possible motion constructed by means of
the exact Sundman inequalities differ from analogous
regions defined according to the simplified Sundman inequality,
both quantitatively and qualitatively.

The stability regions determined by the simplified Sundman
inequality (\ref{D4}) have larger sizes than those calculated
by exact inequality (\ref{D2}). Therefore, the stability obtained by
means of (\ref{D4}) can turn to instability, when using exact
inequality (\ref{D2}).

It is easy to derive by means of the exact
Sundman surfaces that the loss of Sundman stability
for the body $M_3$
can occur only when a certain distance $R$
between the bodies $M_1$ and $M_2$ takes place, so that the "passage"
through the neighborhood of the singular point $L_2$ is open.
It is caused by the fact that the singular points of the Sundman surfaces
are determined by three coordinates $L_i (x_i, y_i, R_i)$
and in the space $xyR$ they lie, generally speaking,
in different planes. This result cannot be established with the aid of inequality
(\ref{D4}), since it does not depend on $R$.

The construction of exact Sundman surfaces allows us
to define the regions of possible motions for any of
the three bodies and for any values of $C$ and $B$.
Using the Sundman surfaces yields, for example, that
with the fulfillment of the stability criterion the body
$M_3$ (it can be any body) for any
time $-\infty< t < \infty$ will be located
at a finite distance from one of the bodies $M_1$ or $M_2$
or at a large distance from these bodies. Qualitatively,
the analogous result is known for the Hill surfaces in the restricted
three-body problem as well.
If the body $M_3$ is located, for example, in the stability region
near $M_1$, then the Sundman surfaces admit the possibility
of retreating of the body $M_2$ to any large
distance from the pair $M_1, M_3$.
For the Hill surfaces this situation is not possible.

By means of the Sundman surfaces it is possible to establish
the stability of only one pair of bodies, and the third body
will be in this case unstable in the Sundman sense. Sundman
surfaces do not establish the simultaneous stability of
three bodies, i.e., guaranteed location of all bodies
in a certain finite region of the space (Lagrange stability),
although these surfaces do not exclude this case.
Sundman instability does not mean that a body will necessarily leave
the neighborhood of another body. The Sundman surfaces do not
allow us to determine if this retreat will actually occur.
This result is analogous to that of Hill stability. The determination
of Sundman stability of the planet satellites of the Solar system
conducted in this study shows the effectiveness of the use of Sundman
surfaces in the coordinate form.

We believe that our results represent a certain
interest for celestial mechanics and for astronomy as a whole.

\newpage
\begin{table}
\caption{Martian satellites. Here $a$ is the semimajor axis of the
satellite orbit, $i$ is the inclination, $e$ is the eccentricity,
$m/M_P$ is the ratio of the satellite mass to the planet mass. }
\vskip2mm \centering
\begin{tabular}{lcccccc}
\hline \qquad Satellite & $a$ & $e$ & $i$ & $m/M_P$ &
\multicolumn{2}{c} {Stability} \\ \hhline{~~~~~--} \quad
\quad & ($km$) &  & ($deg$)& $10^{-8}$ & Hill  & Sundman \\
\hline
 M1 Phobos & 9380    & 0.0151    & 1.1  &   1.6723  & yes & yes \\
 M2 Deimos & 23460    & 0.0002    & 0.9 -- 2.7  & 0.2288 & yes & yes \\
\hline
\end{tabular}
\label{Tab0}
\end{table}

\begin{longtable}{lcccccc}
\caption{The irregular Jovian satellites}\hskip-1cm (the notation in Table
\ref{Tab0}). \\
\hline \centering
\qquad Satellite & $a$
& $i$ & $e$ & $m/M_{P}$ & \multicolumn{2}{c} {Stability}\\
\hhline{~~~~~--}
\quad & $(10^6 km$) & ($deg$) & & $10^{-9}$ &
Hill& Sundman \\
\hline
\qquad\qquad 1&2&3&4&5&6&7\\
\hline
\endfirsthead \multicolumn{7}{c} {Tables 2 continued.} \\ \hline
\qquad\qquad 1&2&3&4&5&6&7\\
\hline \endhead \hline \multicolumn{7}{c} {\textit{Continued
on the next page}} \endfoot \hline \endlastfoot \label{Tab2}
XVIII Themisto  & 7.507 &    43.08 &     0.242 & 3.4889& yes & yes\\
XIII Leda & 11.165 &27.46 & 0.164 & 5.76& yes& yes\\
VI Himalia & 11.461 & 27.50 & 0.162 & 22101.8& yes & yes \\
X Lysithea  &     11.717 &    28.30 & 0.112 & 331.5 & yes & yes \\
VII Elara &         11.741 & 26.63  & 0.217 & 4578.2& yes & yes \\
XLVI   Carpo & 16.989 & 51.4 & 0.430 & 0.3394& yes & yes\\
S/2003 J3 &    18.340  & 143.7 & 0.241 & 0.1263 & no & no \\
S/2003 J12 &    19.002 &    145.8 &     0.376 &     0.0631& yes & yes \\
XXXIV Euporie  &  19.302 &    145.8 &     0.144 & 0.2447 & no & no
\\  S/2003 J18 &  20.700 & 146.5 & 0.119 & 0.2920& no &
no \\ XXXV  Orthosie
&      20.721 &    145.9 &     0.281 &     0.3315& no & no \\
XXXIII  Euanthe  &     20.799 &    148.9 &     0.232 & 0.4341 & no &
no
\\  XXIX    Thyone  &        20.940 & 148.5  & 0.229 & 0.6946 & no &
no \\ S/2003 J16 & 21.000 & 148.6 & 0.270  & 0.1342 & no & no \\
XL  Mneme & 21.069 & 148.6 & 0.227 & 0.3315& no & no \\ XXII
    Harpalyke &    21.105  & 148.6 &     0.226 &0.8367 & no & no \\
XXX     Hermippe  &      21.131 & 150.7 &     0.210 &     1.4919 &
no & no \\ XXVII Praxidike &  21.147  &   149.0  & 0.230 & 2.8495&
no &
no \\ XLII  Thelxinoe &   21.162 & 151.4 & 0.221 & 0.3473 & no & no \\
XXIV    Iocaste  &      21.269 & 149.4  &    0.216
&    1.3971& no & no \\
XII Ananke & 21.276 &    148.9 &     0.244 & 157.9 & no & no \\
S/2003 J15 &    22.000 &    140.8 & 0.110 & 0.1342& no & no \\
S/2003 J4 &     23.258 & 144.9 & 0.204 & 0.0947& no & no \\
L   Herse & 22.000 &    163.7 & 0.190 & 0.2526& no & no\\
S/2003 J9 &     22.442 & 164.5 & 0.269 & 0.0947& no & no \\
S/2003 J19 & 22.800 & 162.9  &    0.334  & 0.1263& no & no \\
XLIII   Arche & 22.931 & 165.0 &     0.259 &     0.2842& no & no
\\ XXXVIII  Pasithee  &       23.096 &    165.1 & 0.267 & 0.1658&
no & no \\ XXI Chaldene  & 23.179 & 165.2 & 0.251 & 0.7499& no &
no \\ XXXVII Kale & 23.217  & 165.0 & 0.260 & 0.2447& no & no\\
XXVI Isonoe & 23.217 &    165.2 & 0.246 &     0.6157& no & no \\
XXXI Aitne & 23.231 &    165.1 &     0.264 & 0.4026 & no & no \\
XXV Erinome &       23.279 & 164.9  & 0.266 & 0.3789& no & no\\
XX Taygete & 23.360 &    165.2 &     0.252 &     1.1445  & no & no \\
XI Carme &          23.404  &   164.9  &    0.253 &694.6 & no & no \\
XXIII Kalyke &      23.583 & 165.2 & 0.245 & 1.5471 & no & no \\
XLVII Eukelade & 23.661 & 165.5 & 0.272 & 0.7104& no & no \\
XLIV    Kallichore  & 24.043  & 165.5 & 0.264 & 0.2289 & no & no \\
S/2003 J5 & 24.084 & 165.0 & 0.210 & 0.9788& no & no \\ S/2003 J10
&    24.250  &   164.1   &   0.214 &     0.0947& no & no \\
XLV Helike &        21.263  &   154.8  &    0.156 & 0.7183& no & no \\
XXXII   Eurydome &   22.865  & 150.3  & 0.276 & 0.4262& no & no \\
XXVIII  Autonoe  & 23.039 & 152.9 & 0.334 & 0.7814 & no & no \\
XXXVI Sponde & 23.487 &    151.0 &     0.312 & 0.2763& no & no \\
VIII Pasiphae &       23.624 & 151.4  &    0.409 & 1578.7 & no & no \\
XIX Megaclite  &     23.806 & 152.8 & 0.421 &2.1312 & no & no \\
IX Sinope & 23.939 & 158.1 & 0.250 & 394.7 & no & no\\ XXXIX
    Hegemone & 23.947  & 155.2  &    0.328 &     0.3394& no & no \\
XLI     Aoede & 23.981 &    158.3  &    0.432   &   0.6473  & no & no \\
S/2003 J23 &    24.055  &   149.2 &     0.309 &     0.0947& no & no \\
XVII Callirrhoe  &    24.102 &    147.1   & 0.283   & 5.3044 & no &
no
\\ XLVIII  Cyllene  & 24.349 & 149.3 & 0.319 & 0.2368& no &
no \\ XLIX  Kore & 24.543 & 145.0  & 0.325 & 0.3947& no & no \\
S/2003 J2 & 28.570 & 151.8 & 0.380 & 0.1500 & no & no \\
\label{Tab1}
\end{longtable}

\begin{table}
\caption{The irregular Saturnian satellites (the notation in Table
\ref{Tab0}).} \vskip2mm \centering
\begin{tabular}{lcccccc}
  \hline
\qquad Satellite & $a$ & $i$ & $e$ & $m/M_P$ & \multicolumn{2}{c}
{Stability}  \\ \hhline{~~~~~--} \quad \quad & ($10^6
km$) & ($deg$) & & $10^{-11}$ & Hill  & Sundman \\
\hline XXIV Kiviuq &        11.111 &      45.71&      0.334& 0.8629&
yes & yes \\ [-2pt] XXII Ijiraq &       11.124 & 46.44& 0.316 &
0.3248& yes &  yes \\ [-2pt] IX Phoebe &
12.944 &      174.8 &     0.164 &     1458.957& yes & yes \\
[-2pt] XX Paaliaq &         15.200 &      45.13 &     0.364 & 2.2728
&  yes & yes \\ [-2pt] XXVII Skathi &       15.541 & 152.6 & 0.270 &
0.0588& yes    & yes\\ [-2pt] XXVI Albiorix &
16.182 &      33.98 &     0.478 &     4.3629  & yes & yes \\
[-2pt] S/2007 S2 &   16.560 & 176.7 & 0.218 & 0.0248& yes & yes \\
[-2pt] XXXVII Bebhionn &
17.119 &      35.01 &     0.469 &     0.0261& yes & yes \\
[-2pt] XXVIII Erriapus &         17.343 & 34.62 &     0.474 & 0.2294&
yes & yes \\ [-2pt] XXIX Siarnaq &     17.531 & 45.56 & 0.295 &
24.1988 & yes & yes \\ [-2pt] XLVII Skoll & 17.665 & 161.2 & 0.464 &
0.0237&    yes & yes \\ [-2pt] LII Tarqeq &
17.920 &      49.86 &     0.107 &     0.0385& yes & yes \\
[-2pt] XXI Tarvos &        17.983 &      33.82 &     0.531 & 0.5455&
yes & yes \\ [-2pt] LI Greip &          18.105 & 172.7 & 0.374 &
0.0158&  yes & no \\ [-2pt] XLIV Hirrokkin
&    18.437 &      151.4 &     0.333 &     0.0965& yes & yes \\
[-2pt] S/2004 S13&     18.450 &      167.4 & 0.273 &     0.0148& yes
& yes \\ [-2pt] S/2004 S17&
18.600 &      166.6 &     0.259 &     0.0082&  yes & yes \\
[-2pt] L Jarnsaxa &    18.600 &      162.9 &     0.192 & 0.0116 &
yes & no \\ [-2pt] XXV Mundilfari &    18.685 & 167.3 & 0.210 &
0.0464 & yes  & no \\ [-2pt] S/2006 S1 &
18.981 &      154.2 &     0.130 &     0.0192& yes & no \\
[-2pt] XXXI  Narvi &           19.007 &      145.8 & 0.431& 0.0340 &
yes  & no \\ [-2pt] XXXVIII Bergelmir &
19.338 &      158.5 &     0.142 &     0.0248& yes & no \\
[-2pt] XXIII Suttungr &         19.459 & 175.8 &     0.114 & 0.0422 &
yes & no \\ [-2pt] S/2004 S12 &    19.650 & 164.0 & 0.401 & 0.0142 &
yes  & no \\ [-2pt] S/2004 S07 &
19.800 &      165.1 &     0.580 & 0.0200 & yes & no \\
[-2pt] XLIII Hati &        19.856 & 165.8 &     0.372 & 0.0185& yes
& no \\ [-2pt] XXXIX Bestla &      20.129 & 145.2 & 0.521 & 0.0432&
yes & no \\ [-2pt] XL Farbauti & 20.390 & 156.4 & 0.206 & 0.0113 &
yes & no \\ [-2pt] XXX Thrymr &
20.474  &     176.0 &     0.470 &     0.8278&    yes & no \\
[-2pt] S/2007 S3 &     20.518 &      177.2 &     0.130  & 0.0119 &
yes & no \\ [-2pt] XXXVI Aegir &        20.735 & 166.7 & 0.252 &
0.0214& yes & no \\ [-2pt] S/2006 S3 &
21.132  &     150.8 &     0.471 &     0.0100  & yes & no \\
[-2pt] XLV Kari &          22.118 &      156.3 & 0.478 & 0.0409 &
yes & no \\ [-2pt] XLI Fenrir  & 22.453  & 164.9 & 0.136 & 0.0095&
yes & no \\ [-2pt] XLVIII Surt & 22.707 & 177.5 & 0.451 & 0.0127&
yes & no \\ [-2pt] XIX Ymir   &
23.040  &     173.1 &     0.335 &     1.3878&   yes & no \\
[-2pt] XLVI Loge &         23.065 &      167.9 &     0.187 & 0.0232
&    yes    & no \\ [-2pt] XLII Fornjot &      25.108 & 170.4 &
0.206 & 0.0211&   yes & no \\ \hline
\end{tabular} \label{Tab2}
\end{table}

\begin{table}
\caption{The irregular Uranus' satellites (the notation in Table
\ref{Tab0}).} \vskip2mm \centering
\begin{tabular}{lcccccc}
  \hline
\qquad Satellite & $a$ & $e$ & $i$ & $m/M_P$ & \multicolumn{2}{c}
{Stability} \\
\hhline{~~~~~--} \quad \quad &  $(10^6 km$) &  &
($deg$)&$10^{-9}$
 & Hill  & Sundman \\
\hline XXII  Francisco   & 4.2760 & 0.1425 & 147.613 & 0.0658& yes
& yes \\  XVI Caliban &  7.1689   & 0.0823    & 139.681 & 8.1305 &
yes & yes \\  XX  Stephano &  7.9424   & 0.1459 & 141.538 & 0.3494 &
yes & yes \\  XXI Trinculo & 8.5040  & 0.2078 & 166.332 & 0.0593 &
yes & yes \\  XVII   Sycorax & 12.2136 & 0.5094 & 152.669 & 46.6790
& yes & yes \\  XXIII Margaret & 14.3450 &   0.7827  & 50.651 &
0.0609  & yes & yes \\
 XVIII  Prospero & 16.1135  & 0.3274    & 146.340 & 1.1306  &
yes & yes \\  XIX Setebos & 18.2052   & 0.4943 & 148.828 & 1.4240 &
yes & yes \\  XXIV    Ferdinand   & 20.9010 & 0.4262 & 167.278 &
0.0874 & yes & no \\
\hline
\end{tabular}
\label{Tab3}
\end{table}

\begin{table}
\caption{The irregular Neptune's satellites (the notation in Table
\ref{Tab0}).} \vskip2mm \centering
\begin{tabular}{lcccccc}
\hline
\qquad Satellite & $a$ & $e$ & $i$ & $m/M_P$ &
\multicolumn{2}{c} {Stability} \\ \hhline{~~~~~--} \quad
\quad & $(10^6
km$) &  & ($deg$)& $10^{-9}$ & Hill  & Sundman \\
\hline
 II Nereid & 5.5134    & 0.7512    & 7.232 &   301.38 & yes & yes \\
 IX Halimede    & 15.728    & 0.5711    & 134.101   & 3.0835 &yes & yes \\
 XI Sao & 22.422    & 0.2931    & 48.511    & 0.6445 &yes. & yes\\
 XII Laomedeia   & 23.571    & 0.4237    & 34.741    & 0.5606 & yes & yes \\
 X Psamathe  & 46.695    & 0.4499    & 137.391   & 0.9244 & yes & no \\
 XIII Neso  & 48.387    & 0.4945    & 132.585   & 1.3423 & yes & no \\
\hline
\end{tabular}
\label{Tab4}
\end{table}

\newpage
\section*{REFERENCES}

Donnison J.R., Williams I.P., 1983, Celest. Mech., 31, 123.

Donnison J.R., 2009, Planet. Space Sci., 57, 771.

Donnison J.R., 2010, Planet. Space Sci., 58, 1169.

Emel'yanov N.V., Arlot J.-E., 2008, Astron. Astrophys., 487, 759.

Emelyanov N.V., Uralskaya V.S., 2011, Solar Syst. Res., 45, 5, 377.

Golubev V.G., 1967, Doklady. Akad. Nauk SSSR, 174, 767.

Golubev V.G., 1968, Sov. Phys. Dokl., 13, 373.

Golubev V.G., Grebenikov E.A., 1985, The three-body problem in
Celestial Mechanics, Moscow University Publisher, Moscow (in
russian).

Hagihara Y., 1952, Japan Academy, 28, Number 2.

Hill G.W., 1878, Am. J. Math., 1, 5.

Li J., Fu Y., Sun Y., 2010, Celest. Mech. Dynam. Astron., 107, 21.

Lukyanov L.G., Shirmin G.I., 2007, Astr. Letters, 33, 550.

Lukyanov L.G., 2011, Astron. Rep., 55, 742.

Marchal C., Saari D., 1975, Celest. Mech., 12, 115.

Marchal C., Bozis G., 1982, Celest. Mech., 26, 311.

Marchal C., 1990, The Three-Body Problem, Elsevier Publisher,
Amsterdam.

Natural Satellites Data Center (NSDC)\\
($http://www.sai.msu.ru/neb/nss/index.htm$)

Proskurin V.F., 1950, Bull. Inst. Theor. Astr., IV, Number 7, 60.

Sundman K.F., 1912, Acta Math., 36, 195.

Szebehely V., Zare K., 1977, Astron. Astrophys, 58, 145.

Walker I.W., Emslie A.G., Roy A.E., 1980, Celest. Mech., 22, 371.

Zare K., 1976, Celest. Mech., 14, 73.

\end{document}